\documentclass[useAMS,usenatbib]{mn2e}
\usepackage{graphicx}

\newcommand{\ltsima} {$\; \buildrel < \over \sim \;$}
\newcommand{\gtsima} {$\; \buildrel > \over \sim \;$}
\newcommand{\cm}{\rm{\, cm}}

\newcommand{\eV}{\rm{\, eV }}
\newcommand{\keV}{\rm{\, keV }}
\newcommand{\MeV}{\rm{\, MeV }}
\newcommand{\GeV}{\rm{\, GeV }}
\newcommand{\TeV}{\rm{\, TeV }}

\newcommand{\flu}{\rm{\, erg \, cm^{-2} \, s^{-1}}}
\newcommand{\llu}{\rm{\, erg \, s^{-1}}}
\newcommand{\beq}{\begin{equation}}
\newcommand{\eeq}{\end{equation}}
\newcommand{\ba}{\begin{array}}
\newcommand{\ea}{\end{array}}

\newcommand{\ei}{\eta_{3}}

\newcommand{\rd}{r_{\gamma,15}}
\newcommand{\ro}{r_{0,8}}

\newcommand{\ee}{\epsilon_{e,-1}}
\newcommand{\eB}{\epsilon_{B,-1}}

\newcommand{\apjl}{\mbox{\it Astrophys. J.}}
\newcommand{\apj}{\mbox{\it Astrophys. J.}}
\newcommand{\apjs}{\mbox{\it Astrophys. J. Supp. Ser.}}
\newcommand{\mnras}{\mbox{\it Mon. Not. R. Astron. Soc.}}

\def \etal{{\it et al.~}}

\title[Connection between thermal and non-thermal emission in GRBs]{The Connection Between Thermal and Non-Thermal Emission in Gamma-ray Bursts: General considerations and GRB090902B as a Case Study} 

\author[A. Pe'er et. al.] {Asaf Pe'er$^{1,2}$, 
Bin-Bin Zhang$^{3}$, 
Felix Ryde$^{4}$, 
Sin{\'e}ad McGlynn$^{4}$, \newauthor 
Bing Zhang$^{3}$, 
Robert D. Preece$^{5}$, 
Chryssa Kouveliotou$^{6}$\\
$^{1}$Space Telescope Science Institute, 3700 San Martin
  Dr., Baltimore, Md, 21218 and \\ Harvard-Smithsonian Center for
  Astrophysics, 60 Garden St., Cambridge, MA, 02138; apeer@cfa.harvard.edu\\
$^{2}$Giacconi Fellow\\
$^{3}$Department of Physics and Astronomy, University of
  Nevada, Las Vegas, NV 89154\\ 
$^{4}$Department of Physics, Royal Institute of Technology,
AlbaNova, SE-106 91 Stockholm, Sweden\\
$^{5}$Center for Space Plasma and Aeronomic Research
  (CSPAR), University of Alabama in Huntsville, Huntsville, AL 35899,
  USA\\
$^{6}$Space Science Office, VP62, NASA/Marshall Space
  Flight Center, Huntsville, AL 35812, USA
}

\begin{document}

\date{Accepted... Received...; in original form \today}

\pagerange{\pageref{firstpage}--\pageref{lastpage}} \pubyear{2011}

\maketitle

\label{firstpage}
 
\begin{abstract}
  Photospheric (thermal) emission is inherent to the gamma-ray burst
  (GRB) ``fireball'' model. We show here, that inclusion of this
  component in the analysis of the GRB prompt emission phase naturally
  explains some of the prompt GRB spectra seen by the {\it Fermi}
  satellite over its entire energy band. The sub-MeV peak is explained
  as multi-color black body emission, and the high energy tail,
  extending up to the \GeV band, results from roughly similar
  contributions of synchrotron emission, synchrotron self Compton
  (SSC) and Comptonization of the thermal photons by energetic
  electrons originating after dissipation of the kinetic energy above
  the photosphere. We show how this analysis method results in a
  complete, self consistent picture of the physical conditions at both
  emission sites of the thermal and non-thermal radiation. We study
  the connection between the thermal and non-thermal parts of the
  spectrum, and show how the values of the free model parameters are
  deduced from the data. We demonstrate our analysis method on
  GRB090902B: We deduce a Lorentz factor in the range $920 \leq \eta
  \leq 1070$, photospheric radius $r_{ph} \simeq 7.2 - 8.4 \times
  10^{11}$~cm and dissipation radius $r_\gamma \geq 3.5 - 4.1 \times
  10^{15}$~cm. By comparison to afterglow data, we deduce that a large
  fraction, $\epsilon_d \approx 85\% - 95\%$ of the kinetic energy is
  dissipated, and that large fraction, $\sim$~equipartition of this
  energy is carried by the electrons and the magnetic field.  This
  high value of $\epsilon_d$ questions the ``internal shock'' scenario
  as the main energy dissipation mechanism for this GRB.
\end{abstract}


\section{Introduction}
\label{sec:intro}

Although extensively studied for nearly two decades, the origin of the
prompt emission of gamma-ray bursts (GRBs) is still puzzling. Up until
these days, GRB prompt emission spectra are often modelled
as a smoothly broken power-law, which is referred to as the ``Band''
function \citep{Band93, Preece98a, Preece00, Kaneko06, Kaneko08,
  Abdo09b}. While the ``Band'' function often provides very good fits
to the observed spectra over a limited energy range, it suffers two
crucial drawbacks. In several bursts seen by the {\it Fermi}
satellite, a high energy tail extending up to tens of \GeV was seen
(e.g., GRB090510, \cite{Ackerman10}; or GRB090902B,
\cite{Abdo09}). The first drawback is that this tail requires more
than the ``Band'' function on its own to have an acceptable fit.
However, the most severe drawback is that the ``Band'' function, being
empirical in nature, does not provide any information about the
emission mechanism that produces the prompt radiation.

A common interpretation of the observed GRB spectrum is that it
results from synchrotron emission, accompanied perhaps by
synchrotron-self Compton (SSC) emission at high energies \citep{RM94,
  SP97, PL98, GuG03, PW04, GZ07}. The emission follows the dissipation of a
kinetic energy. The prevalent dissipation models involve either
internal shocks \citep{RM94, SP97b}, magnetic energy dissipation in
Poynting dominated outflows \citep{Thompson94, SDD01, GS05, ZY11}, or
collisional heating \citep{Bel10}.  These models have two main
advantages. First, they can explain the complex lightcurves often seen
(albeit with very little predictive power). Second, they naturally
account for the non-thermal spectrum.

In spite of these advantages, in recent years evidence is accumulating
for serious difficulties in these models. A well known deficiency of
the internal shock scenario is the low efficiency of energy conversion
\citep{MMM95, KPS97, DM98, LGC99, GSW01, MZ09}, resulting from the
fact that only the energy associated with the differential motion
between the expanding ejecta shells after the GRB explosion can be
dissipated, and that this energy is much lower than the energy
associated with the bulk motion. This is in contrast to the
observations which show high efficiency in $\gamma$-ray emission, of
the order of tens of percent \citep{Zhang07, NFP09}.  A second
drawback is that optically thin synchrotron and SSC emission cannot
account for the steepness of the low energy spectral slopes
\citep{Crider97,Preece98b,Preece02,GCG03} \citep[although part of the
observed steepening may be accounted for when SSC in the Klein Nishina
limit is considered; see][]{DBD09, Bosnjak09}.  Third, the high energy
spectral slope varies significantly from burst to burst
\citep{Preece98a, Kaneko06}, with some bursts showing very steep high
energy spectral slopes. This is in contradiction to the expected
spectral slope of synchrotron emission from a power law distribution
of electrons, which is expected to produce a fairly flat spectrum,
$\nu F_\nu \propto \nu^0$.  Finally, the fitting is often made to the
{\it time integrated} spectrum. Analysis of {\it time resolved}
spectra done by \citet{CLP98} and \citet{GCG03}, has shown that
neither the synchrotron nor the SSC models can explain the time
resolved low energy spectral slopes.

These well known difficulties of the synchrotron emission model have
motivated the study of alternative scenarios. These include, among
others, quasi-thermal Comptonization \citep{GC99}, Compton drag
\citep{ZSP91, Shemi94, LGCR00}, jitter radiation \citep{Med00},
Compton scattering of synchrotron self absorbed photons \citep{PM00,
  SP04} and synchrotron emission from a decaying magnetic field
\citep{PZ06}.

An alternative model, which is arguably the most natural one, is a
radiative contribution from the photosphere \citep{EL00, MR00, MRRZ02,
  DM02, RM05, Ghirlanda+07, Bel10, Ioka10, MNA11}.  Indeed, a
photospheric emission is a natural outcome of the ``fireball''
model. At small radii, close to the inner engine, the optical depth is
huge, $\tau \sim 10^{15}$ \citep[see, e.g., ][ for a
review]{Piran05}. As a result, photons cannot escape, but are advected
with the flow until they decouple at the photosphere. The huge value
of the optical depth implies that regardless of the initial emitted
spectrum, if photons are emitted deep enough in the flow, the spectrum
emerging at the photosphere is black-body (a Wien spectrum may be
obtained if the number of photons is conserved). In fact, in the
classical ``fireball'' model, photons serve as mediators for energy
conversion (the explosion energy is converted into kinetic energy of
the relativistically expanding plasma jet by scattering between
photons and leptons, which in turn convert the energy to the baryons
via Coulomb collisions). Therefore, the appearance of thermal photons
as part of the observed spectrum is not only expected, but is in fact
{\it required}.

Apart from being an inherent part of the ``fireball'' model, one of
the key advantages of the idea that at least part of the observed
spectrum originates from the photosphere is that it provides a relief
for the efficiency problem of the internal shock model \citep{PMR05,
  RP09, LMB09}. This is because if indeed part of the photons that we
see have photospheric origin, than the total energy seen in photons is
higher than the energy released by the internal dissipation: only the
non-thermal part of the spectrum is required to originate from energy
dissipation above the photosphere. Thus, the dissipated energy may be
smaller than the remaining kinetic energy, even though the total
energy seen in the photon component is comparable to the kinetic
energy.  A second great advantage of this idea is that the
Rayleigh-Jeans part of the thermal spectrum (or modification of it)
can naturally account for low energy spectral slopes much steeper than
those allowed by the synchrotron theory, and are hence consistent with
observations.

The observed GRB spectra are therefore expected to be hybrid, i.e.,
containing both thermal and non-thermal parts.  These two parts are
connected in a non-trivial way. As the non-thermal part originates
from energy dissipation (one or more) occurring above the photosphere,
it is naturally delayed with respect to the thermal photons.  A
complete decomposition of the thermal and non-thermal parts of the
spectrum is in fact nearly impossible. Thermal photons serve as seed
photons to Compton scattering by the energetic electrons, thereby
affecting the non-thermal part as well. Even if the dissipation occurs
high above the photosphere, thermal photons significantly contribute
to the cooling of the electrons \citep{PMR05}, therefore affecting the
spectrum emitted by these electrons. As was shown by \citet{PMR06},
under these conditions a flat energy spectrum ($\nu F_\nu \propto
\nu^0$), resulting from multiple Compton scattering of the thermal
photons, is naturally obtained for a large range of parameters. A
similar conclusion was drawn in the case of a magnetized outflow under
the assumption of slow heating due to continuous reconnection at all radii
\citep{G06}. These predictions were shown to be in qualitatively very
good agreement with {\it Fermi} results \citep{TWM11}.

The non-trivial connection between the thermal and non-thermal
components, combined with the fact that both components vary in time,
make it difficult to identify the thermal component. A breakthrough in
identifying this component was made by \citet{Ryde04, Ryde05}, who
looked at {\it time resolved} spectra, thereby allowing an
identification of the temporal evolution of the temperature.
Repeating a similar analysis on a fairly large sample of bursts, it
was shown by \citet{RP09} that the thermal component not only could be
identified, but that both the temperature and flux of this component
show repetitive temporal behavior at late times: $F_{BB}^{ob} \propto
t^{-2}$, and $T^{ob} \propto t^{-2/3}$. This behavior was found to be
in very good agreement with the theoretical predictions
\citep{Peer08}, thereby providing an independent indication for the
correct identification of the thermal emission component. It should be
pointed out here, that the thermal emission is expected to appear as
gray-body emission, composed of multi-color black body spectra. This
results from the fact that at a given time interval, an observer sees
photons emitted from different radii and different angles, hence
undergoing different Doppler shifts. A full analysis shows that the
resulting low energy spectrum (below the observed spectral peak,
typically seen at sub \MeV) is a power law \citep{PR10}.

Once the thermal component is identified, it is relatively easy to use
its properties to deduce the dynamics of the outflow. As opposed to
the non-thermal part, whose emission radius is uncertain (there may be
multiple emission radii), the emission radius of the thermal photons
is defined to be at the photosphere.  Thus, by studying the properties
(thermal flux and temperature) of the thermal component, under the
assumption of constant outflow velocity, it is possible to deduce the
photospheric radius and the Lorentz factor of the flow
\citep{PRWMR07}. We note that this method is independent, and is
complementary to the opacity argument method commonly used to
constrain the Lorentz factor of the flow at the emission radius of the
high-energy photons \citep{Svensson87, KP91, WL95, LS01}. It has two
main advantages to the opacity argument method. First, it provides a
direct measurement of the Lorentz factor, rather than a lower
limit. Second, it is independent of measurements of the variability
time, which is highly uncertain.

While in the past a clear identification of a thermal component was
difficult, the situation dramatically changed with the broad band
spectral coverage enabled with the launch of {\it Fermi}. Out of 14
GRBs detected by the {LAT} until January 2010, 3 show clear evidence
for a distinctive high energy ($> \MeV$ and up to the \GeV range)
spectral component, and another 8 show marginal evidence for such a
component \citep{Granot10}. Thus, it is natural to deduce that the low
energy photons (below and at the sub \MeV peak of the flux)
have a different origin than the higher energy (LAT) photons.

Out of the 3 bursts that show clear evidence for a distinct high
energy spectral component, GRB090902B may be the easiest to analyse,
for two reasons. First, the low energy part of the spectrum (the
spectral peak) is very narrow, and both the low energy and high energy
spectral slopes are very steep. Thus, any attempts of fitting the
spectrum using the standard synchrotron-SSC model are rejected. On the
other hand, the spectrum is easily fitted with a (multi-color) black
body spectrum, plus an additional power law \citep{Ryde+10,
  Zhang+10}. Thus, to date, this burst is unique by the fact that a
thermal component is so clearly pronounced in its spectrum.  Second,
the high energy power law component is spectrally distinctive from the
low energy one. Using the opacity argument, the detection of a $33.4
\GeV$ photon associated with this burst \citep{Abdo09} implies that
the emission radius of this photon must be much greater than the
photosphere (see below).  These two facts make this burst ideal for
demonstrating how separation of the spectrum into thermal and
non-thermal components enables one to deduce the physical conditions
of the outflow at the emission sites of both the thermal and
non-thermal photons.  Moreover, as we will show below, one can use
measurements of the non-thermal part, to remove some of the
uncertainties that exist in measurements of the thermal component
alone.

This paper is organized as follows. In section \S\ref{sec:general}, we
provide a general discussion on the properties of the thermal - non
thermal model. We show how one can combine the hydrodynamic
information derived by studying the properties of the thermal
component with the constraints given by measuring the properties of
the non-thermal part of the spectrum, to provide a comprehensive
picture of the physical parameters at both emission sites.  In
\S\ref{sec:fits} we demonstrate the use of our method by fitting the
prompt emission spectrum of GRB090902B, and deducing its physical
properties. We summarize and conclude in \S\ref{sec:summary}.

\section{General properties of a thermal - non-thermal model}
\label{sec:general}

\subsection{Temperature, luminosity and constraints on the value of $\epsilon_e$}
\label{sec:temp_luminosity}

We consider a fireball wind of total luminosity $L$, expanding from an
initial radius $r_0$ (which, for the sake of argument, can be assumed
to be a few times the last stable orbit around the central black hole, or the
sonic radius; in any case, it is a few times the Schwarzschield radius
around a non-rotating black hole; see further discussion at the end of
section \ref{sec:thermal}). The initial black-body temperature
at $r_0$ is $T_0 = (L/4\pi r_0^2 c a)^{1/4}$, where $c$ is the speed
of light and $a$ is the radiation constant. As the optically thick
(adiabatic) wind expands, the baryon bulk Lorentz factor increases as
$\Gamma \propto r$, and the comoving temperature drops as $T' \propto
r^{-1}$ \citep[see, e.g., ][for a comprehensive review]{Mes06}.
\footnote{From here on, quantities measured in the comoving frame are
  primed, while unprimed quantities are in the observer frame.}  As
long as the wind remains optically thick, the acceleration continues
until the plasma reaches the saturation radius, $r_s = \eta r_0$,
above which $\Gamma$ coasts to a value equal to the dimensionless
entropy, $\eta \equiv L/{\dot M} c^2$. Here, ${\dot M}$ is the mass
outflow rate.

The photospheric radius is the radius above which the flow becomes
optically thin to scattering by the baryon-related
electrons. Depending on the values of the free model parameters
($\eta, L,$ and $r_0$), this radius can be above or below the
saturation radius \citep[see][]{MRRZ02}. For the parameter values 
characterizing GRBs (see below), the photospheric radius is above the
saturation radius, and is given by \citep{ANP91, Peer08}
\beq
r_{ph} = {L \sigma_T \over 8 \pi m_p \eta^3 c^3} = 5.8 \times 10^{11}
L_{54} \ei^{-3} \cm. 
\label{eq:r_ph}
\eeq
Here and below, $\sigma_T$ is the Thomson cross section, $m_p$ is the proton
mass, and we use the convention $Q= 10^x Q_x$ in cgs units. The high
values of the luminosity and the Lorentz factor chosen for the
demonstration in equation \ref{eq:r_ph}, are for ease of
comparison with the {\it Fermi} results of GRB090902B (see \S\ref{sec:fits}
below).  

Above the saturation radius, adiabatic energy losses (in the absence of
dissipation) cause the temperature to drop as $T = T_0
(r_s/r)^{-2/3}$. The observed temperature of photons emitted at the
photosphere is therefore
\beq
T^{ob} = T_0 \left( {r_{ph} \over r_s} \right)^{-2/3} = 3.6 \times
  10^5 (1+z)^{-1} \, L_{54}^{-5/12} \ei^{8/3} \ro^{1/6} \eV,
\label{eq:temp}
\eeq
where $z$ is the redshift. Note the very strong dependence of the
observed temperature on the asymptotic value of the Lorentz factor,
$\eta$: for high $\eta$, high values of the temperature are expected.

The observed photospheric thermal luminosity drops above the
saturation radius as $L_{Th}(r) = (L/2) (r/r_s)^{-2/3}$, the greater
part of the energy being in a kinetic form, $L_k \sim L/2$
\citep{MR00, RM05}\footnote{Note that in fact at $r>r_{ph}$ the
  kinetic luminosity is slightly higher due to energy conversion from
  the photons above the saturation radius.  While the full treatment
  will be given below, in the content of equation \ref{eq:luminosity}
  this has little effect, and is omitted for clarity.}. The
non-thermal part of the spectrum results from dissipation of the
kinetic energy. Part of the dissipated energy goes into accelerating
electrons, that radiate the non-thermal spectrum. Denoting by
$\epsilon_d$ the fraction of kinetic energy that is dissipated and by
$\epsilon_e$ the fraction of dissipated energy that is converted to
energetic electrons, one obtains an {\it upper limit} on the ratio of
non-thermal to thermal luminosity in the spectrum:
\beq
{L_{NT}^{ob} \over L_{Th}^{ob}} \leq \epsilon_d \epsilon_e \left( {r_{ph} \over r_s}
\right)^{2/3} = 0.33 \, L_{54}^{2/3} \ei^{-8/3} \ro^{-2/3} \epsilon_{d,0} \ee,
\label{eq:luminosity}
\eeq
where the non-equality results from the fact that the electrons do not
necessarily radiate 100\% of their energy.  Equation
\ref{eq:luminosity} implies an interesting result: {\it the higher the
  Lorentz factor of the flow is, the more pronounced its thermal
  luminosity is expected to be}. This result may be very significant
given recent {\it Fermi}-LAT data, which show evidence for high values
of the Lorentz factors in several bursts.
We further point out  that as $\eta$ increases, the saturation radius
$r_s$ increases, while the photospheric radius decreases. Thus, at
high enough value of $\eta = \eta^* \equiv (L \sigma_T /4 \pi m_p c^3
r_0)^{1/4}$,  $r_{ph} = r_s$, and the ratio
$L_{NT}^{ob}/ L_{Th}^{ob}$ saturates to $\epsilon_e \epsilon_d$
\citep[see][for further details]{MRRZ02, PRWMR07}.
Higher value of $\eta > \eta^*$ does not change this result.

\subsection{Initial expansion radius, photospheric radius and Lorentz factor}
\label{sec:Gamma}

As was shown by \citet{PRWMR07}, the outflow parameters, in particular
the Lorentz factor $\eta$, the initial expansion radius, $r_0$ and the
photospheric radius, $r_{ph}$ can be inferred directly from studying
the thermal component alone (for bursts with known redshift). For
completeness of the analysis, we briefly repeat here the main
arguments given by \citet{PRWMR07}.

The observed temperature of thermal photons emitted from the
photosphere is $T^{ob} \simeq 1.48 \eta T'(r_{ph})/(1+z)$
\footnote{For photons emitted along the line of sight, the Doppler shift
  is $\mathcal{D}(\theta=0) = 2 \eta$. The numerical factor  $1.48$
  results from angular integration.}.  
Due to Lorentz aberration, the ratio $(F_{Th}^{ob}/\sigma
{T^{ob}}^4)^{1/2}$ is proportional to the photospheric radius divided
by the Lorentz factor, 
\beq
\mathcal{R} \equiv \left( {F_{Th}^{ob} \over \sigma {T^{ob}}^4}
\right)^{1/2} = \xi {(1+z)^2 \over d_L} {r_{ph} \over \eta}.
\label{eq:R}
\eeq 
Here, $\sigma$ is Stefan's constant, $\xi$ is a geometrical factor of
order unity, $d_L$ is the luminosity distance
and $F_{Th}^{ob} = L_{Th}^{ob}/ 4 \pi d_L^2$ is the observed thermal
flux. Using the parametric dependence of $r_{ph}$ from equation
\ref{eq:r_ph} in equation \ref{eq:R}, one obtains the asymptotic
Lorentz factor,
\beq
\eta = \left[ \xi (1+z)^2 d_L {F_{Th}^{ob} \sigma_T \over 2 m_p
    c^3 \mathcal{R}} \right]^{1/4} \left({L \over L_{Th}^{ob}}
\right)^{1/4}.
\label{eq:eta}
\eeq
Combining  this result with the equation for the observed thermal
flux, $L_{Th}^{ob} = (L/2) (r_{ph}/r_s)^{-2/3}$, one obtains the initial
expansion radius,
\beq
r_0 = \left({d_L \mathcal{R} \over \xi (1+z)^2}\right)
\left({L \over 2 L_{Th}^{ob}}\right)^{-3/2},
\label{eq:r0}
\eeq
and the photospheric radius,
\beq
r_{ph} = \left[ {d_L^5 F_{Th}^{ob} \sigma_T \mathcal{R}^3 \over
  \xi^3 (1+z)^6 2 m_p c^3} \right]^{1/4} \left( {L \over
  L_{Th}^{ob}} \right)^{1/4}.
\label{eq:r_ph2}
\eeq

The values of $\eta$, $r_0$ and $r_{ph}$ are thus fully determined by
the observed quantities of the thermal emission, up to the uncertainty
in the luminosity ratio $L/L_{Th}^{ob} \geq 1$. Furthermore, equations
\ref{eq:eta} and \ref{eq:r_ph2} imply that the values of $\eta$ and
$r_{ph}$ are not very sensitive to the uncertainty in this ratio.

Constraining the ratio of the total luminosity released in the
explosion to the luminosity emitted as thermal photons,
$L/L_{Th}^{ob}$, can most easily be done if an independent measurement
of the kinetic energy exists. Such measurements are provided by
studying the emission during the afterglow phase \citep{WG99, Frail01,
  PK01, FW01, BFK03, BKF03, NFP09}, which provides good estimates of
the kinetic energy remaining after the prompt emission
phase. Fortunately, such measurements become ubiquitous, and are
available for GRB090902B \citep{Cenko11}; see further discussion in
\S\ref{sec:fits} below.

Even if afterglow measurements do not exist, the luminosity ratio
$L/L_{Th}^{ob}$ can still be constrained indirectly, in three
independent methods. First, by fitting the non-thermal part of the
spectrum, one obtains a constraint on $\epsilon_d \epsilon_e
(L/L_{Th}^{ob})$ (see eq. \ref{eq:luminosity}). Since $\epsilon_d
\epsilon_e < 1$, a lower limit on $L/L_{Th}^{ob}$ is obtained. Second,
variability time measurements (if they exist) can constrain the
initial expansion radius, since $\delta t \geq r_0/c$, which, in turn,
provides an upper limit on the ratio $L/L_{Th}^{ob}$ via equation
\ref{eq:r0}.  And finally, a high value of $L_{Th}^{ob}$ implies that
the luminosity ratio should not be high, in order to avoid an energy
crisis.

\subsection{Constraint on the emission radius of the non-thermal
  photons}
\label{sec:opacity}

Observations of high energy (\gtsima $10 \GeV$) photons by
{\it Fermi}, are commonly used in the literature to constrain the
Lorentz factors of GRB outflows, using the opacity argument. We point
out though, that the constraints set in the literature are often based
on an additional assumption, that is that the emission radius of the
high energy photons, $r_\gamma$ is connected to the Lorentz factor via
$r_\gamma = 2 \eta^2 c \delta t$, where $\delta t$ is the variability
time of the inner engine activity. This assumption, while true in the
internal shocks model scenario, has two main drawbacks. First, it
assumes an a-priori knowledge of the variability in the Lorentz
factor, namely $\Delta \eta \simeq \eta$ (this assumption translates
into the numerical coefficient); and second, it relies on an assumed
knowledge of the physical variability time, $\delta t$, which is
difficult to be measured accurately.

A different approach was suggested by \citet{ZP09}: by releasing the
requirement $r_\gamma = 2 \eta^2 c \delta t$, it was shown that the
opacity argument can be used to provide general constraints in the
$r_\gamma - \eta$ space. While this method does not provide directly
the value of $\eta$ (or of $r_\gamma$), its main advantage is that it
is not sensitive to the uncertainties mentioned above. Here, we take
the arguments presented by \citet{ZP09} one step forward. We first use
the analysis of the thermal component to estimate the Lorentz factor
$\eta$. At the second step, we use the constraints found by the
opacity argument in the $r_\gamma - \eta$ plane to deduce a lower limit
on the emission radius of the non-thermal photons, $r_\gamma$.

The calculation is performed as follows. The cross section for pair
production of photon with energy $\varepsilon_1$ is the highest for
interactions with photons of energy $\varepsilon_2 = (m_e
c^2)^2/\varepsilon_1$. Therefore, considering the Lorentz boosting, a
photon observed with energy $\varepsilon_{\max}^{ob} = 10 \,
\varepsilon_{\max,10}^{ob} \GeV$ interacts mainly with photons having
energies
\beq 
\varepsilon_1^{ob} \leq {(m_e c^2)^2 \eta^2 \over \varepsilon_{\max}
  (1 + z)^2} = {26 \over (1+z)^2} \, \ei^2
{(\varepsilon_{\max,10}^{ob})}^{-1} \MeV.
\label{eq:eps_1}
\eeq 
This energy is about two orders of magnitude higher than the energy of the
thermal photons (see eq. \ref{eq:temp}). We therefore do not expect
the thermal photons to play a significant role in constraining the
emission radius of the most energetic photons seen by the {\it Fermi}-LAT.

The observed spectrum above a few \MeV is often modeled by a single
power law, $dN^{ob}/dt^{ob} dAd\varepsilon^{ob} = f_0
{\varepsilon^{ob}}^{-\alpha}$.  For such a spectral fluence, the
optical depth for pair production can be written as \citep{KP91, WL95,
  LS01, ZP09}
\beq
\tau_{\gamma \gamma} = { <\sigma> d_L^2 \over r_\gamma^2 (1+z)^2} {f_0
  \Delta t_{\GeV}^{ob} \over \alpha -1} \left[ {(m_e c^2)^2 \over
    \varepsilon^{ob}_{\max}} \right]^{1- \alpha} \left( {\eta \over 1
    + z } \right)^{2 - 2\alpha}.
\label{eq:tau_gg}
\eeq 
Here, $<\sigma> $ is the cross section averaged over all angles; for
flat energy spectrum ($\alpha=2$), one obtains $<\sigma> \approx
\sigma_T/8$ \citep{Svensson87,GZ08}\footnote{Note that this value is
  about twice the value presented in \citet{LS01}.}. Further note that
$\Delta t_{\GeV}^{ob}$ in equation \ref{eq:tau_gg} represents the observed
time bin during which high energy photons are seen, and thus does not
necessarily correspond directly to the uncertain physical variability
time.\footnote{E.g., $\Delta t_{\GeV}^{ob}$ could correspond to the
  integrated time over several distinct events.}

For a flat energy spectrum ($\alpha=2$) observed between
$\varepsilon_{\min}$ and $\varepsilon_{\max}$\footnote{This assumption
  is taken here as it provides a good first order approximation to the
  observed high energy spectrum. A more accurate calculation considering
  the exact power law index will be used in \S\ref{sec:fits}.}, the
observed fluence is related to the (non thermal) luminosity via $f_0 =
L_{NT}^{ob}/4 \pi d_L^2 \log(\varepsilon_{\max}/\varepsilon_{\min})$.
The requirement that the optical depth to pair production of the
highest energy photon seen is smaller than unity, is translated into a
lower limit on the emission radius,
\beq
\ba{lcl}
r_\gamma & \geq & \left[{<\sigma> L_{NT}^{ob} \Delta t_{\GeV}^{ob} \varepsilon_{\max}^{ob}
    \over 4 \pi \log(\varepsilon_{\max}/\varepsilon_{\min}) }
\right]^{1/2} {1 \over \eta m_e c^2} \nonumber \\
&= &3 \times 10^{15} \, L_{54}^{1/2} \,
{\Delta t_{\GeV, 0}^{ob}}^{1/2} \, {\varepsilon_{\max,10}^{ob}}^{1/2} \,
\ei^{-1} \, \cm, 
\ea
\label{eq:r_gamma}
\eeq
where $\log(\varepsilon_{\max}/\varepsilon_{\min}) \approx 20$ was
taken. For parameters characterizing GRBs seen by the LAT, the
emission radius of the non-thermal photons is about 3 - 4 orders of
magnitude larger than the photospheric radius (eq. \ref{eq:r_ph}),
indicating that the observed spectrum must be emitted from (at least)
two separate regions.

\subsection{ The observed non-thermal spectrum:  additional constraints on the
free model parameters}
\label{sec:spectrum}

The dissipation at $r_\gamma$, regardless of its exact nature,
produces a population of energetic electrons. The energetic electrons
emit the non-thermal part of the spectrum, by radiating their
energy. There are three main radiative mechanisms responsible for the
non-thermal emission: synchrotron emission, synchrotron-self Compton
(SSC) emission and Comptonization of the thermal photons. 

In order to estimate the relative contributions of the different
emission mechanisms to the observed spectrum, we proceed in the
following way.

\subsubsection{Electron energy loss by Comptonization of the thermal
  photons}
\label{sec:energy_density}

The thermal photons serve as seed photons for Compton scattering by
the energetic electrons.  Their existence therefore contributes to the
high energy part of the non-thermal spectrum.

The power emitted by Comptonization of the thermal photons (in the
Thomson regime), relative to the power emitted as synchrotron
radiation, is given by the ratio of the energy density of the thermal
photon field to the energy density in the magnetic field.  At the
photosphere, the (comoving) energy density of the thermal photons is
$u_{Th}(r_{ph}) = a T'(r_{ph})^4 = (L/8\pi r_s^2 c
\eta^2)(r_{ph}/r_s)^{-8/3}$.  Since above the saturation radius the
energy density of the photon field drops as $u_{Th} \propto r^{-2}$,
at the dissipation radius $r_\gamma$ it is equal to $u_{Th} (r_\gamma)
= (L/8\pi r_\gamma^2 \eta^2 c)(r_{ph}/r_s)^{-2/3}$. The energy density
in the magnetic field assumes a fraction $\epsilon_B$ of the comoving
energy density, $L/8\pi r_\gamma^2 \eta^2 c$. One therefore concludes
that the ratio of the energy densities in the thermal photon and
magnetic field is given by
\beq
{\widetilde Y} \equiv {u_{Th} \over u_B} = {1 \over \epsilon_B}
\left({r_{ph} \over r_s}\right)^{-2/3} = 3 \, L_{54}^{-2/3} \, \ei^{8/3}
\ro^{2/3} \, \eB^{-1}
\label{eq:Y}
\eeq
It is thus clear that for parameters characterizing GRBs, the role
played by Comptonization of thermal photons is, at the least,
comparable to the role played by the synchrotron emission as a source
of energy loss of the energetic electrons.

\subsubsection {Synchrotron spectrum}
\label{sec:sync}

The dissipation process is expected to produce a power law
distribution of energetic electrons with power law index $p$\gtsima$
2.0$, above a characteristic Lorentz factor $\gamma_m \simeq
\epsilon_e (m_p/m_e) = 184 \ee$. \footnote{If the dissipation results
  from shock waves crossing, this equation implies a mildly
  relativistic shock Lorentz factor, $\Gamma_s -1 \approx 1$, and is
  thus consistent with the internal shocks model.}  Comparison of the
cooling time to the dynamical time implies that the entire electron
population is in the fast cooling regime: the cooling time is equal to
the dynamical time for electrons having Lorentz factor $\gamma_c = (3
\pi m_e c^3 \eta^3 r_\gamma)/[\sigma_T \epsilon_B L (1 + Y +
{\widetilde Y})] = 3.5 \, (1+Y+{\widetilde Y})^{-1} \, L_{54}^{-1} \,
\ei^3 \, \rd \, \eB^{-1}$. Here, $Y$ has its usual meaning as the
ratio of SSC to synchrotron radiated power, $Y \equiv
P_{SSC}/P_{syn}$.

The peak of the synchrotron emission, $\varepsilon_m^{ob} = (3/2)
\hbar \eta/(1+z) (q B \gamma_m^2 / m_e c) = 1.5 \, (1+z)^{-1} L_{54}^{1/2}
\eB^{1/2} \rd^{-1} \ee^2 \keV$, is below the threshold energy of the
{\it Fermi}-GBM detector. Here, $q$ is the electron charge and $B$ is
the magnetic field at the dissipation radius.  At the other end of the
energy spectrum, comparison of the cooling time to the acceleration
time provides an estimate of the maximum Lorentz factor of the
accelerated electrons (assuming high efficiency in the acceleration
process), $\gamma_{\max} = [6 \pi q / \sigma_T B (1 + Y + {\widetilde
  Y})]^{1/2}$. Synchrotron photons emitted by these electrons are
expected at energies $\varepsilon_{\max,syn}^{ob} = 240 \, (1+z)^{-1}
\, \ei \, (1 + Y + {\widetilde Y})^{-1} \GeV$, above the threshold
energy for pair production, and above the maximum photon energy seen
so far by {\it Fermi}. These results imply that synchrotron emission
is expected to contribute to the spectrum at the entire spectral range
covered by {\it Fermi}.

The fast cooling of the electrons imply that: (I) virtually all of the
dissipated energy given to the electrons is radiated; and (II) for
power law index $p \approx 2.0$, a flat energy spectrum ($\nu F_\nu
\propto \nu^0$) is expected from synchrotron emission over the entire
energy range covered by {\it Fermi}. Note though that a high energy
cutoff resulting from pair production can limit the maximum observed
photon energy to values lower than $\varepsilon_{\max,syn}^{ob}$ (see
\S\ref{sec:opacity} above).

\subsubsection{Comptonization}
\label{sec:Compton}

There are two sources of Comptonized spectrum: Comptonization of the
thermal photons, and SSC. At low energies, below the thermal peak,
Comptonization is not a significant source of photons. Hence, the
spectrum below the thermal peak is dominated by synchrotron
emission. However, above the thermal peak, the three emission
mechanisms- synchrotron, SSC and Comptonization of the thermal photons
contribute in parts to the spectrum. As shown in equation \ref{eq:Y}
and is further discussed below, the relative contributions of the
different emission mechanisms are of the same order of magnitude (in
other words, both $Y$ and $\widetilde Y$ are of the order unity). As a
result, it is difficult to determine a single dominant emission
mechanism at high energies. A consequence of this, is that the
observed spectral index at high energies (at the LAT band, above the
thermal peak) cannot be directly related to a power law index of the
energetic electrons.

{\it Comptonization of the thermal emission}.  Since $(\gamma_m
T')/m_e c^2 = 0.13 \, L_{54}^{-5/12} \ei^{5/3} \ro^{1/6} \ee < 1$,
Comptonization of the thermal emission is in the Thomson limit.  The
Comptonized thermal photon spectrum has characteristic breaks similar
to the SSC spectrum. Since the electrons are in the fast cooling
regime, the Comptonized spectrum is expected to rise above $\gamma_c^2
T^{ob}$ \gtsima $T^{ob}$ up to a peak at $\varepsilon_{Th}^{IC, ob} =
(4/3) \gamma_m^2 (2.8 T^{ob}) \approx 50\, (1+z)^{-1} \, L_{54}^{-5/12}
\ei^{8/3} \ro^{1/6} \ee^2 \, \GeV$, roughly as $\nu F_\nu \propto
\nu^{1/2}$ \citep{SE01}. At the highest energies, photons annihilate
by producing pairs, a phenomenon which can explain the lack of
detection of the $50 \GeV$ photons so far.

{\it SSC}. The ratio of SSC to synchrotron luminosity is given by the
parameter $Y$.  For $\epsilon_e \gg \epsilon_B$, the value of $Y$ can
be approximated as $Y \approx (\epsilon_e/\epsilon_B)^{1/2}$
\citep{SE01}.\footnote{Note that if $Y=\tilde{Y}$, one obtains $Y
  \simeq (\epsilon_B/2\epsilon_e)^{1/2}$.}
 A significant high-energy non-thermal part, as is seen
by the {\it Fermi}-LAT, implies (via eq. \ref{eq:luminosity}), high
value of $\epsilon_e$, close to equipartition. However, value of
$\epsilon_B > 10^{-2}$ as is inferred in many GRBs (see below),
guarantees a value of $Y$ of a few at most.  

The SSC spectrum rises as $\nu F_\nu \propto \nu^{1/2}$ below the peak
of the SSC emission, which is expected at $\varepsilon_m^{IC, ob} = 2
\gamma_m^2 \varepsilon_m^{ob} \approx 100 (1+z)^{-1} L_{54}^{1/2}
\eB^{1/2} \ee^4 \rd^{-1} \MeV$. At higher energies, the spectral shape
follows a similar power law as the synchrotron spectrum, i.e., a flat
energy spectrum is expected for $p \approx 2$.

The rise parts of both the thermal Comptonization and the SSC emission
are independent of the power law index of the accelerated
electrons. For power law index $p$\gtsima $2.0$ a flat, or slightly
decaying synchrotron (energy) spectrum is expected at the entire {\it
  Fermi} energy range (see \S\ref{sec:sync} above). The combined
effects of the flat synchrotron emission with the rise of the
Comptonized spectrum therefore results in a mild increase in the total
observed spectrum at high energies, above the thermal peak (see
\S\ref{sec:fits} below). Demonstration of spectral decomposition into
its basic physical ingredients is presented in
\S\ref{sec:decomposition}.

The rising of the SSC component below $\sim 100 \MeV$, combined with
the fact that the synchrotron spectrum is expected to be flat (or
slightly inverted) and that $Y$ is not expected to be much larger than
a few, imply that at low energies (below the thermal peak),
Comptonization is not expected to play a significant role. The main
emission mechanism below the thermal peak is therefore synchrotron
emission. Since the {\it Fermi}-GBM detection range is above
$\epsilon_m^{ob}$, the spectrum below the thermal peak is expected to
be sensitive to the power law index of the accelerated electrons,
$p$. Thus, measurement of the flux at low energies can provide an
indication for both the values of $\epsilon_B$ and of $p$.

Finally, we point out that the analysis carried in this section holds
only as long as $r_\gamma \gg r_{ph}$. For $r_\gamma$ \gtsima $r_{ph}$,
the effect of Comptonization is much more complicated due to the fact
that the electrons cooling time is $\propto r_\gamma$. Therefore, for
small dissipation radius, the cooling time is much faster than the
dynamical time for all electron energies, and $\gamma_c $\gtsima
$1$. For such rapid cooling, additional physical phenomena, that are
not considered here, become important. Direct Compton scattering of
the energetic photons provide the main source of heating, resulting in
a quasi steady state distribution of mildly-relativistic electrons
\citep{PMR05}. Close to the photosphere, multiple Compton scattering
by electrons in this quasi steady state produces a flat energy spectrum
for a large region of parameter space \citep{PMR06}.

\section{Demonstration of the analysis method: GRB090902B as a
  concrete example}
\label{sec:fits}

The bright, long GRB090902B, which is one of the brightest bursts
observed by LAT to date \citep{Abdo09}, provides an excellent example
for demonstrating our analysis method. This is because of two very
pronounced properties of its prompt emission spectrum. First, time
resolved spectral analysis reveals a significant power law component
in the LAT data (emission was observed up to 30 \GeV), which is
clearly distinct from the usual ``Band'' function used by
\citet{Abdo09} to fit the spectrum in the sub-MeV range
\citep{Ryde+10, Zhang+10}. Moreover, the
fact that the high energy photons were delayed with respect to the
photons at the sub MeV peak indicates a different origin. Second, the
``Band'' function used in fitting the sub MeV peak is extremely steep
on both sides (low energy spectral slope $n(\varepsilon) \propto
\varepsilon^\alpha$, with $\alpha \approx -0.5 .. 0.3$, and high energy
spectral slope $\beta \approx -3 .. -5$), resulting in an unusually
narrow peak.  The steep spectral slopes seen in the sub-MeV range make
it impossible to fit this spectrum with a model that contains only
synchrotron and SSC.

On the other hand, both properties of the prompt spectrum fit
perfectly into the framework suggested here: first, the sub-MeV peak
is naturally modeled with the thermal emission component. While a
single black body provides a sufficient fit, and is used to deduce the
values of $\eta, r_0$ and $r_{ph}$, better fits are obtained with
multi-color black body, as expected from a theoretical point of view
\citep{PR10}.  Second, the non-thermal part (the high energy power law
which extends to lower energies), can be easily explained by a
combination of synchrotron, SSC and Comptonization of the thermal
photons. Moreover, by doing so, we deduce the physical properties in
the emission sites of both the thermal and non-thermal components,
hence we obtain a comprehensive physical picture of the properties of
this burst.

In the original analysis, \citet{Abdo09} separated the observed prompt
emission into several time bins. The most restrictive constraints on
the emission radius of the non-thermal photons arise in time interval
(c), $9.6 - 13.0$~s from the trigger, since in this time interval an
$11.16 \GeV$ photon was observed. We therefore focus our analysis on
this time interval.

\subsection{Analysis of the thermal component: Lorentz factor,
  photospheric radius and dissipation radius} 
\label{sec:thermal}

In order to deduce the value of the Lorentz factor by using the method
presented in \S\ref{sec:Gamma}, one needs to fit the sub-MeV peak with
a single black body spectrum, and study its properties (temperature
and thermal flux). While this can be done for the data in time
interval (c), since the properties of the outflow vary on a shorter
time scale (variability time of \ltsima $0.1$~s was observed),
smearing of the black body spectrum is expected. The shorter the time
interval used for the fits, the higher the quality of the black body
fits obtained \citep{Ryde+10, Zhang+10}. On the other hand, shorter
time intervals result in lower quality of the high energy data, which
is sparse.

We thus use time interval as short as possible to fit the narrow peak
with a single black body, in order to deduce the hydrodynamics of the
flow, and in particular obtain the Lorentz factor. We then use the
longer time interval (the full time interval (c)) to study the
properties of the non-thermal part of the spectrum. For the long time
interval, we fit the spectral peak with a multicolor black body,
which, as was shown by \citet{Ryde+10}, provides better fits to the
peak. Clearly, by choosing to fit the longer time interval during which
the values of the parameters vary, we lose an accuracy in the
fits. However, we stress here, that our purpose in this paper is {\it
  not} to provide the best statistical fits (in terms of $\chi^2$) to
the data. Instead, our goal here is to prove that one can obtain an
{\it acceptable} fits to the data in the {\it entire Fermi energy
  band}, in the sense that the curves plotted all fall within the $\pm
1 \sigma$ error bars of the data points. By doing so, we show that we
are able to provide an acceptable, complete physical interpretation
to the observed data, although, necessarily, we are not able to
capture many second-order effects.

For the shorter time interval we use the results presented by
\citet{Ryde+10}, who showed that the narrow time interval, $11.008 -
11.392$~s from the trigger, during which the $11.16 \GeV$ photon is
seen, provides sufficient data to analyze the sub-MeV spectral
peak. During this time interval, the sub-MeV peak can be fitted with
a single black body, with observed temperature $T^{ob} = 168$~keV, and
observed thermal flux $F_{Th}^{ob} = 1.96 \times 10^{-5} \flu$.
Equation \ref{eq:R} thus implies a ratio $\mathcal{R} = 1.55\times
10^{-19} \cm$. At redshift $z=1.822$ \citep{Cucchiara09}, the
(isotropic-equivalent) thermal luminosity is $L_{Th}^{ob} = 4 \pi
d_L^2 F_{Th}^{ob} = 4.6 \times 10^{53} \llu$.  Using $\xi = 1.06$ in
equations \ref{eq:eta}, \ref{eq:r0} and \ref{eq:r_ph2}
\citep{PRWMR07}, one obtains $\eta = 764 \, (L/L_{Th}^{ob})^{1/4}$,
$r_{ph} = 6.0 \times 10^{11} \, (L/L_{Th}^{ob})^{1/4} \cm$ and $r_0 =
7.9 \times 10^8 \, (L/2 L_{Th}^{ob})^{-3/2} \cm$.

Estimating the ratio $(L/L_{Th}^{ob})$ is done in the following way.
Due to the rapid cooling of the electrons, nearly 100\% of the energy
that is converted to the energetic electrons in the dissipation
process is radiated in the form of non-thermal photons, that is
$L_{NT}^{ob} \simeq \epsilon_e \epsilon_d L_k$. Here, $L_k$ is the
energy available in kinetic form above the photosphere.  By fitting
the non-thermal part of the spectrum, one obtains $L_{NT}^{ob} \simeq
0.9 L_{Th}^{ob}$, or $\epsilon_e \epsilon_d \sim 0.9
(L_{Th}^{ob}/L_k)$ (see Figure \ref{fig:ee}). These fits are best done
using a numerical simulation, since part of the dissipated energy is
released outside the observed energy range of {\it Fermi}.  After the
main dissipation, the remaining kinetic luminosity, $L_{AG} = L_k ( 1
-\epsilon_d)$ is the available luminosity for the afterglow emission
phase. As was found by \citet{Cenko11}, the energy release during the
afterglow phase is $\sim$~five times less than the energy release
during the prompt emission phase. However, since the luminosity in
time interval (c) is about twice the average luminosity during the
prompt phase, we estimate the luminosity (thermal + non-thermal) in
time interval (c) to be about ten times higher than the luminosity
during the afterglow phase,
\beq
{L_{AG} \over L_{Th}^{ob} + L_{NT}^{ob}} = {L_k(1-\epsilon_d)
\over L_{Th}^{ob} + L_k \epsilon_e \epsilon_d} \approx {1 \over 10}.
\label{eq:L_AG}
\eeq 
 Using $\epsilon_e \epsilon_d \sim 0.9 (L_{Th}^{ob}/L_k)$, one obtains
 the relation
\beq
\epsilon_d = { 9 \over 1.9 \epsilon_e + 9}
\label{eq:epsilon_d}
\eeq
This result immediately implies very high dissipation efficiency, since
for any value of $\epsilon_e$, $\epsilon_d $\gtsima $0.83$. 

At the photosphere, the thermal luminosity is equal to $L_{Th}^{ob} =
(L/2) (r_{ph}/r_s)^{-2/3}$, and the kinetic luminosity is equal to
$L_k = (L/2)[2-(r_{ph}/r_s)^{-2/3}]$. At larger radii, the kinetic
luminosity is assumed unchanged, up until the dissipation
radius. Using the observed relation $L_{NT}^{ob}/L_{Th}^{ob} =
\epsilon_d \epsilon_e L_k / L_{Th}^{ob} \simeq 0.9$ and the results
obtained in equation \ref{eq:epsilon_d}, one obtains the
luminosity ratio , 
\beq
{L \over L_{Th}^{ob}} \simeq { (1.19 \epsilon_e + 0.9) \over  \epsilon_e}.
\label{eq:N}
\eeq

For equipartition value, $\epsilon_e = 0.33$, one obtains
$L/L_{Th}^{ob}= 3.9$, which imply $\eta = 1070$ (and $\epsilon_d =
0.94$). However, we consider this value of the Lorentz factor as an
upper limit, due to two reasons: first, this result implies very high
total GRB luminosity, $L = 1.8\times 10^{54} \llu$; and second, using
equation \ref{eq:r0}, it implies an initial expansion radius $r_0
\simeq 3.0 \times 10^8 \cm$, which translates into very short
variability time, $\delta t = r_0/c = 10$~ms. Lower value of
$\epsilon_e$ results in higher Lorentz factor, but also leads to
significantly higher total luminosity, significantly lower variability
time and nearly 100\% dissipation efficiency, which we consider
unlikely.

Using the extreme value $\epsilon_e = 1$,
one obtains $L/L_{Th}^{ob} = 2.09$, $\eta = 920$, $\epsilon_d = 0.83$
and  $r_0 \simeq 7.4 \times 10^8 \cm$, which
translates into physical variability time $\delta t \approx 25$~ms.
We thus conclude, that the asymptotic value of the Lorentz factor is
in the range $920 \leq \eta \leq 1070$, and the value of $\epsilon_e$
is at or slightly above equipartition. We further deduce that the photospheric
radius is $r_{ph} \simeq 7.2 - 8.4 \times 10^{11} \cm$, and that the
dissipation efficiency is $\epsilon_d \approx 85\% - 95\%$. Using these
values of $\eta$ in equation \ref{eq:r_gamma}, leads to the conclusion
that the emission radius of the non-thermal photons is constrained,
$r_\gamma \geq 3.5 - 4.1 \times 10^{15} \cm$.

The derived value of $r_0 \simeq 3.0 - 7.5 \times 10^8$~cm implies
$r_0 \simeq 33-80 \,r_{ISCO,10}$, where $r_{ISCO,10} = 6 G M/c^2$ is the
inner most stable circular orbit (ISCO) of a non-rotating 10-solar
mass black hole. As the exact mass of the progenitor of GRB090902B is
unknown, we can deduce that $r_0$ is of the order of few tens of the
ISCO radius. According to the theory adopted here, $r_0$ marks the
initial expansion radius. As the theory of jet acceleration is not
fully developed yet, the values obtained may thus be used to
constrain models of jet acceleration to this scale. Alternatively, we
note that this value is very close to the value obtained in several
numerical models \citep[e.g.,][]{Aloy+00,ZWM03}. While in these models
the jets are assumed to be produced closer to the ISCO radius, the
acceleration begins only at larger radii due to the fact that at
smaller radii the jet is not well collimated.

\subsection{Numerical calculations: further constraints on the free
  model parameters}
\label{sec:results}

As discussed in \S\ref{sec:spectrum} above, following the dissipation
process, simple analytical descriptions are insufficient to describe
the spectrum, due to the roughly similar contributions from the
different physical phenomena (synchrotron, SSC and thermal
Comptonization), as well as the cutoff resulting from pair production.
Therefore, in order to derive the spectral dependence on the
different values of the free parameters, as well as confirm the
analytical calculations presented above, we calculate numerically the
photon and particle energy distribution for the different
scenarios. In our calculations, we use the time-dependent numerical
code presented in \citet{PW05}. This code solved self-consistently the
kinetic equations that determine the temporal evolution of $e^{\pm}$
and photons, describing cyclo-synchrotron emission, synchrotron self
absorption, direct and inverse Compton scattering, pair production and
annihilation and the evolution of high energy cascade.

This code has two great advantages, which make it ideal in the study
of the prompt spectra. First, it has a unique integrator, that enables
solving the rate equations that govern the time evolution of the
particles and photons energy distribution over the entire energy
range. The code is able to calculate simultaneously processes
happening over more than 15 orders of magnitude in time and energy
ranges, thereby covering the entire spectral range, from radio up to
the \TeV band. The second advantage of the code is a full treatment of
the various physical process, including the full cross sections (e.g.,
Klein Nishina effect is inherently taken into account; or that for
mildly relativistic electrons, the full cyclo-synchrotron emission
spectrum is calculated; etc.).

In modeling the spectrum in time interval (c), the sub MeV peak is best
described as a multicolor black body spectrum, with varying amplitudes
\citep{Ryde+10}: $F_{Th}^{ob}(\nu ; T_{\max}) =
\int^{T_{\max}}_{T_{\min}} (dA/dT) B_\nu (T) dT$, where $ B_\nu (T) =
(2h/c^2) \nu^3/(e^{h \nu/T}-1)$ is Planck function. The amplitude
$A(T)$ is temperature dependent, $A(T) = A(T_{\max})
(T_{\max}/T)^{4-q}$, and is normalized such that the total flux is
equal to the observed flux at the spectral peak, $F_{Th}^{ob} = 1.82
\times 10^{-5} \flu$. The normalization constant $q$ and $T_{\max}$
are determined by fitting the spectrum, $T_{\max} = 328 \keV$ and
$q=1.49$. $T_{\min}$ cannot be determined, and its exact value is
unimportant for the fits, as long as $T_{\min} \ll T_{\max}$.

While the best value for $q$ is found by fitting the data, we note
that this multi-color description of the thermal part of the spectrum
is inherent to emission from relativistically expanding plasma.  At
any given instance, an observer sees simultaneously photons emitted
from a range of radii and angles \citep{Peer08}. The Doppler boosting
of photons emitted at high angles to the line of sight is smaller than
that of photons emitted on the line of sight. Therefore, a pure Planck
function in the comoving frame is inevitably observed as multi-color
black body.

In the framework developed here, the value of $q$ is related to the
spectral index via $F_{\nu} \propto \nu^{q-1}$. This can be seen by
noting that $dA/dT \propto T^{q-5}$, and hence $F_{th}(\nu) \propto
\int dT (dA/dT) B_\nu(T) \propto \nu^{q-1}$; the last equality is
easily obtained by replacing the integrand from $T$ to $z = h
\nu/T$. Thus, the value of $q$ found by fitting the data implies
spectral index $F_{\nu} \propto \nu^{0.49}$. This index is
significantly softer than the index expected for pure black body
($F_{\nu} \propto \nu^2$), or from the scenario considered by
\citet{Bel10}, of fixed comoving temperature, in which $F_{\nu} \propto
\nu^{1.4}$.  On the other hand, this index is harder than the expected
index at late times, $F_{\nu} \propto \nu^0$ \citep{PR10}. At these
times, off-axis emission dominates the spectra. When the spectrum is
dominated by on-axis emission, in the spherically symmetric scenario
it is expected to be $F_{\nu} \propto \nu^1 $ [Lundman et. al., 2011,
in prep.]. The fitted spectral index is thus in between these
values. We find this result encouraging, given that (I) here the
fitted spectrum is integrated over a finite time interval, hence we
average over spectra obtained at different times, and (II) the
theories are developed for the 'pure' case of constant outflow Lorentz
factor. Hence, the results obtained are in good agreement with the
theoretical expectations.\footnote{Additional broadening may result
 from sub-photospheric dissipation, see \citet{Ryde+11}}

In producing the spectrum, we assume that at radius $r_{\gamma}$, a
fraction $\epsilon_d$ of the kinetic energy is being dissipated (by an
unspecified dissipation process). The energetic electrons, which
assume a power law distribution with power law index $p$, carry a
fraction $\epsilon_e$ of the dissipated energy, and the magnetic field
carries a fraction $\epsilon_B$ of this energy.  The multicolor black
body spectra serve as background spectra for all the various processes
(mainly Compton scattering by the energetic electrons, but also other
processes such as, e.g., pair production). The code
tracks the evolution of the spectrum, during the dynamical time. Here,
we present the results at the end of time interval (c), i.e., we
assume that the processes take place during an observed time of 3.4
seconds.

\subsection{Numerical results}
\label{sec:numerical_results}

\begin{figure}
\begin{center}
\rotatebox{0}{\resizebox{!}{60mm}{\includegraphics{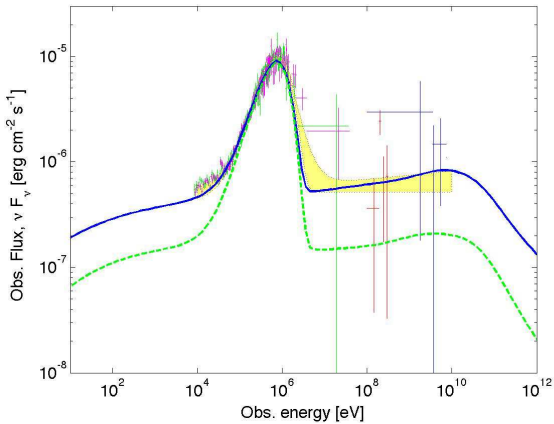}}}
\caption{The dependence of the non-thermal flux on the fraction of
  energy given to the electrons, $\epsilon_e$. The data of GRB090902B
  are from the NaI (0,1), BGO(0,1) and LAT (back and front) detectors
  at time interval (c), 9.6 - 13.0 seconds after the GBM trigger. The
  light yellow shaded area show the $\pm 1 \sigma$ fit to the data at
  this time interval, made by \citet{Abdo09}. Note that this area is
  calculated by \citet{Abdo09} by fitting a ``Band'' plus a single power
  law spectra, hence the apparent discrepancy between the high energy
  (LAT) data points and the shaded area.  In our work, the parameter
  values chosen are: dissipation radius $r_\gamma = 10^{17} \cm$, bulk
  motion Lorentz factor $\Gamma = 910$, power law index of the
  accelerated electrons $p=2.0$, GRB luminosity $L = 9.2 \times
  10^{53} \llu$ and fraction of dissipated kinetic energy $\epsilon_d
  = 0.9$. Shown are the simulation results for $\epsilon_e = 0.5$
  (blue, solid line), and $\epsilon_e = 0.17$ (dashed, green
  line). The $\sim \MeV$ peak is fitted with a (multi-color) black body
  spectrum; the non-thermal flux is linear in the value of
  $\epsilon_e$.  }
\label{fig:ee}
\end{center}
\end{figure}

\begin{figure}
\begin{center}
\rotatebox{0}{\resizebox{!}{60mm}{\includegraphics{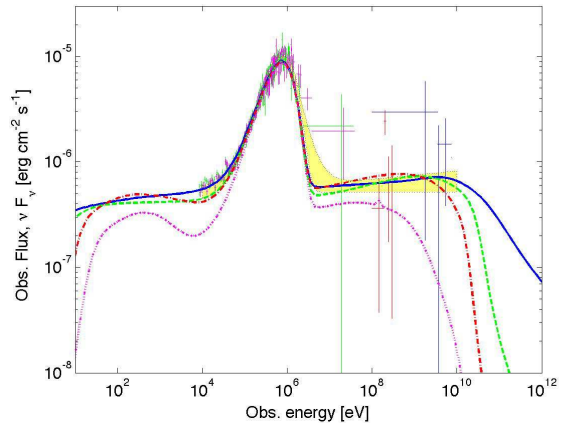}}}
\caption{The dependence of the non-thermal flux on the dissipation
  radius, $r_{\gamma}$. We show the numerical results for dissipation
  occurring at $r_\gamma = 10^{17} \cm$ (solid, blue), $10^{16} \cm$
  (dashed, green), $10^{15.5} \cm$ (dash-dotted, red) and $10^{15}
  \cm$ (dotted, purple), superimposed on the {\it Fermi} data and the
  $\pm 1 \sigma$ ``Band'' function fit to the data (light yellow
  shaded area). Values of $\epsilon_B = 0.33$ and $\epsilon_e = 0.5$
  were chosen, apart from the fit for $r_{\gamma} = 10^{16} \cm$,
  where $\epsilon_e = 0.4$ is chosen.  All the other parameters are
  the same as in Figure 1.  Below $10^{15.5} \cm$, pair production
  limits the maximum observed energy of photons to \ltsima $\GeV$,
  and is thus inconsistent with the LAT observation of $11.16 \GeV$
  photon at this time interval. }
\label{fig:r}
\end{center}
\end{figure}

\begin{figure}
\begin{center}
\rotatebox{0}{\resizebox{!}{60mm}{\includegraphics{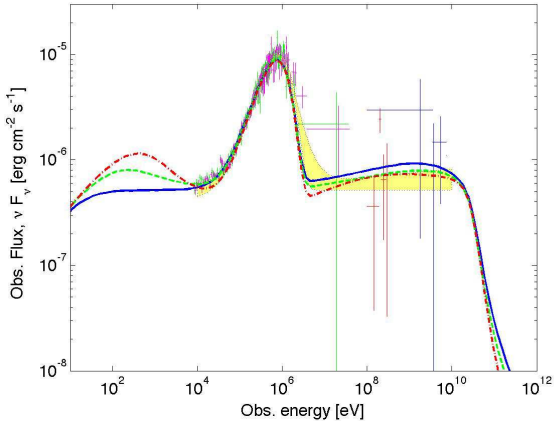}}}
\caption{The dependence of the non-thermal flux on the power law index
  of the accelerated electrons. On top of the {\it Fermi} data, shown
  are the numerical results for $p=2.0$ (solid, blue), $p=2.2$
  (dashed, green) and $p=2.5$ (dash-dotted, red). Dissipation
  radius $r_\gamma = 10^{16} \cm$, $\epsilon_e = 0.5$, $\epsilon_B =
  0.33$ and all other parameter values same as in Figure 1 are
  chosen. The high energy spectrum is nearly insensitive to the exact
  value of $p$ in the range considered, $2.0 - 2.5$. However, the low
  energy part (below the thermal peak) may provide indication for $2.0
  \leq p \leq 2.2$.}
\label{fig:pl}
\end{center}
\end{figure}

\begin{figure}
\begin{center}
\rotatebox{0}{\resizebox{!}{60mm}{\includegraphics{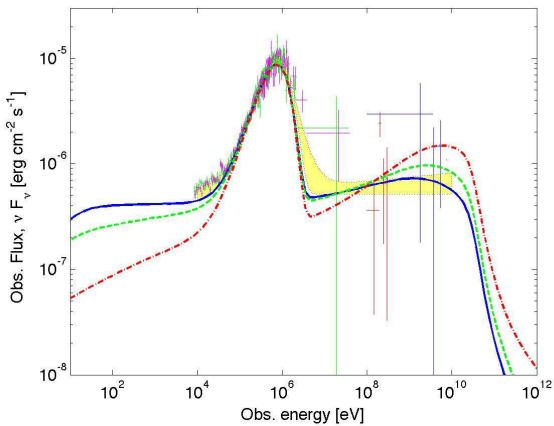}}}
\caption{The dependence of the non-thermal flux on the fraction of
  energy carried by the magnetic field, $\epsilon_B$. Shown are the
  numerical results for equipartition ($\epsilon_B = 0.33$; solid,
  blue), $\epsilon_B = 0.1$ (dashed, green) and $\epsilon_B = 0.01$
  (dash-dotted, red). Dissipation radius $r_{\gamma} = 10^{16} \cm$,
  $\epsilon_e = 0.4$ and all other parameter values same as in figure
  1 are chosen. While the effect of $\epsilon_B$ on the high energy
  spectrum is minor, the low energy flux (below the thermal peak)
  necessitates high value of $\epsilon_B$, close to equipartition. }
\label{fig:eB}
\end{center}
\end{figure}

The numerical fits to the spectrum of GRB09092B at time interval (c),
9.6-13.0 seconds after the trigger, are presented in Figures
\ref{fig:ee} -- \ref{fig:eB}.\footnote{These can be compared to the
  spectra presented in Figure
  3 in \citet{Abdo09}, although note that in this work the spectrum is
  presented at a different time interval. However, the spectral shape and the
  main spectral features (such as the ratio of the flux at the peak to
  the flux at $\sim  10\keV$) are similar.}
In Figure \ref{fig:ee}, we demonstrate
the linear dependence of the non-thermal flux on the value of
$\epsilon_e$. For the fits shown in this figure, we chose parameter
values that fulfill the requirements in the previous sections and are
typical for GRBs. Thus, we chose large dissipation radius, $r_\gamma =
10^{17}$~cm, strong magnetic field, $\epsilon_B = 0.1$, and electron
power law index $p=2.0$.  The Lorentz factor was chosen to be $\Gamma
= 910$, which implies $L/L_{Th}^{ob} = 2.0$, or $L = 9.2 \times
10^{53} \llu$. These values imply, via equation \ref{eq:N},
$\epsilon_e = 0.5$ and $\epsilon_d = 0.9$. The fit to the {\it Fermi}
(GBM + LAT) data at time interval (c) appears as the solid (blue) line
in Figure \ref{fig:ee}. The fit shows the combines spectrum resulting
from both the thermal peak (at $\sim \MeV$), and the nearly flat energy
spectrum ($\nu F_{\nu} \propto \nu^0$) resulting from synchrotron
emission from electrons in the fast cooling regime.
 We further added a scenario in which $\epsilon_e$ is
three times smaller ($\epsilon_e = 0.17$, dashed green line), which
demonstrates the linear dependence of the non-thermal flux on the value
of $\epsilon_e$. We point out that although a power law index $p=2.0$
was chosen, the combined effects of flat ($\nu F_{\nu} \propto \nu^0$)
synchrotron spectrum and rising ($\nu F_{\nu} \propto \nu^{1/2}$)
Comptonization spectrum, lead to a slight increase in the high energy
spectral slope, which is consistent with the slope seen with {\it Fermi}.

We further note that the fit to the Wien part of the thermal component
falls slightly below the $\pm 1 \sigma$ error bars of the data (the
shaded, yellow areas in the figures).  This discrepancy can be easily
understood as due to smearing of the data: the data presented in the
figures is averaged over several seconds, during which the properties
of the outflow (such as the Lorentz factor) slightly vary, while in
the numerical fit we assume steady values of the physical
parameters. Variation in the parameters values inevitably lead to
smearing of the signal, which is translated to a high energy decay
which is somewhat shallower than the exponential cutoff of the thermal
spectrum considered by the fits. Indeed, detailed analyses of time
resolved spectra done by \citet{Ryde+10} and \citet{Zhang+10} show
that as the time interval considered becomes shorter, the exponential
decay above the thermal peak becomes more and more pronounced, hence
the (multi color) black body function used in fitting the peak becomes
better the shorter the time bin is.  Nonetheless, as explained above,
we chose here to fit the data in the entire time interval (c), since
reducing the time interval results in poor quality of the high energy
part of the data (the non-thermal part). We therefore find it
appropriate to use the fits presented, as they serve the main goal of
this paper: to demonstrate that the physically motivated, hybrid
(thermal + non-thermal) model provides acceptable fits, which,
moreover, enable a good estimate of the physical conditions at both
emission sites.

In Figure \ref{fig:r}, we consider different dissipation radii:
$r_{\gamma} = 10^{17} \cm$ (solid, blue), $10^{16} \cm$ (dashed,
green), $10^{15.5} \cm$ (dash-dotted, red) and $10^{15} \cm$ (dotted,
purple). As the numerical code considers the full cross section for
pair production, the numerical results are more accurate than the
analytical approximations presented in \S\ref{sec:opacity}, and can be
used to validate them.  The results presented in Figure \ref{fig:r}
indeed confirm the main conclusion obtained analytically, that is that
the observation of the $11.16 \GeV$ photon necessitates the
dissipation radius to be above $10^{15.5} \cm$.

At larger radii, the high energy non-thermal part of the spectrum is
not very sensitive to the exact dissipation radius. As shown in Figure
\ref{fig:r}, for dissipation radii $r_{\gamma} \geq 10^{15.5} \cm$ it
is possible to obtain numerical results which are within $\pm 1
\sigma$ errors of the empirical ``Band'' fit. In order to achieve
this, high value of $\epsilon_B$ and a slight tuning of the
value of $\epsilon_e$ is required.  Thus, for $r_{\gamma} = 10^{16}
\cm$, a value of $\epsilon_e = 0.4$ was chosen, while for the other
fits in this figure, $\epsilon_e = 0.5$ was found adequate.

The numerical results show that the dependence of the pair production
cutoff at high energies on $r_{\gamma}$ is weaker than the analytical
approximation presented in equation \ref{eq:r_gamma}. The main reason
for this is the contribution from the thermal photons, which is
neglected in the derivation of equation \ref{eq:r_gamma}.  We can
therefore conclude, that further constraints on the dissipation radius
can not be obtained directly from the prompt spectrum, without
additional assumptions. Finally, we point out that the small bump
obtained in the scenario of $r_\gamma = 10^{15} \cm$ at $\Gamma m_e
c^2/ (1+z) \simeq 150 \MeV$, results from pair annihilation process,
which is more pronounced at small dissipation radii due to the more
rapid production of $e^{\pm}$ pairs. Such a small bump is difficult to
be observed, and indeed has not been observed so far. Moreover, since
the Lorentz factor of the flow likely varies during the time interval
considered here, it is expected to be smeared.

In Figures \ref{fig:pl} and \ref{fig:eB} we examine the dependence of
the spectra on the uncertain values of the power law index of the
energetic electrons and the fraction of dissipated energy carried by
the magnetic field. In Figure \ref{fig:pl}, we consider three values
of the power law index: $p=2.0$ (solid, blue), $p=2.2$ (dashed, green)
and $p=2.5$ (dash-dotted, red). As explained in section
\ref{sec:Compton} above, and is further demonstrated in appendix
\ref{sec:decomposition}, the high energy part of the spectrum (above
the thermal peak) is governed by Compton scattering which is in the
rising part of the spectrum (below $\epsilon_m^{IC}$), and therefore
the spectrum is not sensitive to the exact power law index of the
accelerated electrons. We thus conclude that for GRB090902B-type
bursts, for which the thermal peak is pronounced, \footnote{Some other
  bursts, such as, e.g., GRB080916C show much less pronounced thermal
  peak, hence Comptonization may play a sub-dominant role for
  GRB080916c-type bursts.} {\it
  observations at high energies cannot be used to constrain the power
  law index of the accelerated electrons}. On the other hand, the low
energy part of the spectrum (below the thermal peak) is dominated by
synchrotron emission, and is thus sensitive to the power law index of
the electrons. Unfortunately, most of the effect is expected below the
threshold energy of the {\it Fermi}- GBM detector, and therefore only
weak observational constraints exist. We can therefore conclude that a
power law index in the range $2.0 \leq p \leq 2.2$ is consistent with
the data, and even the higher value of $p=2.5$ can be consistent,
however for such high power law index a somewhat fine tuning of the
other model parameters (in particular, very high value of
$\epsilon_B$, close to equipartition) is required.

Examination of the spectral dependence on the value of $\epsilon_B$ is
presented in Figure \ref{fig:eB}. The three fits presented in this
figure, equipartition ($\epsilon_B = 0.33$; solid, blue), $\epsilon_B
= 0.1$ (dashed, green) and $\epsilon_B = 0.01$ (dash-dotted, red) show
that the value of $\epsilon_B$ cannot be too far below equipartition:
low value of $\epsilon_B$ results in a too low flux at low energies
(below the thermal peak), which is inconsistent with the
observation. It also leads to a more pronounced Compton peak, which is
seen at the high energies. We can thus conclude that value of
$\epsilon_B$ close to equipartition is needed to be consistent with the
observations. This high value justifies the need for numerical
analysis, since although $\epsilon_e$ is very high, both $Y$ and
$\tilde{Y}$ are close to unity.

\section{Summary and discussion}
\label{sec:summary}

In this paper, we considered the effect of thermal emission on the
observed GRB prompt emission spectrum. Being a natural outcome of the
GRB fireball model, thermal emission is an inherent part of the prompt
emission seen. As we showed in \S\ref{sec:temp_luminosity} (eq.
\ref{eq:luminosity}), it is expected to be more pronounced for bursts
with higher Lorentz factor. As we show here, {\it the inclusion of the
  thermal emission in the calculation of the prompt emission spectra,
  enables one to obtain a complete, self consistent physical model of
  the prompt emission spectrum seen over the entire FERMI energy
  range}. This is in contrast to the ``Band'' function fits, which do
not carry any physical interpretation, and, in addition, require extra
models to be able to fit the spectrum at high energies.

According to our model, the sub-MeV peak often seen is interpreted as
being composed of multi-color black body emission from the
photosphere. The non-thermal, high energy part seen in several bursts
by the {LAT} instrument is interpreted as resulting from combined
emission of synchrotron, SSC and Comptonization of the thermal
photons, following an episode of energy dissipation that occurs at
large radius above the photosphere, $r_\gamma > r_{ph}$. Observations of high energy
photons can be used to constrain the dissipation radius
(eq. \ref{eq:r_gamma}), which is clearly above the photospheric radius
(eq. \ref{eq:r_ph}). Moreover, as we showed in \S\ref{sec:spectrum},
the relative contributions of synchrotron emission, SSC and
Comptonization of the thermal photons, denoted by the parameters $Y$
and $\tilde{Y}$, are expected to be of the same order of
magnitude. Hence, the three emission mechanisms have roughly similar
contributions to the high energy (above the thermal peak) part of the
spectrum. This fact makes it difficult to directly determine the
power law index of the accelerated electrons from measurements of the
high energy spectral slope.

The separation made here between thermal and non-thermal emission,
makes it possible to deduce the values of the free model
parameters. First, by analyzing the photospheric part, one can deduce
the value of the Lorentz factor, the initial expansion radius and the
photospheric radius (see \S\ref{sec:Gamma}). Then, by analyzing the
non-thermal part, one can constrain the dissipation radius, $r_\gamma$
(eq. \ref{eq:r_gamma}), and place constraints on the power law index
of the accelerated electrons, $p$, and the strength of the magnetic
field, $\epsilon_B$. By combining the fluxes of the thermal and
non-thermal parts, one can further constrain the combined fractions of
dissipated kinetic energy ($\epsilon_d$) that is received by the
energetic electrons ($\epsilon_e$). Further separation of the values
of these variables is more tricky, but can be done with the help of
afterglow observations (see \S\ref{sec:thermal}, eqs. \ref{eq:L_AG} -
\ref{eq:N}).

We demonstrated our analysis method on the bright, long GRB090902B.
This burst is ideal for our demonstration purposes, because of the
clear separation between the thermal and non thermal components seen,
and the very pronounced thermal peak: both above and below the sub-MeV
peak seen in this burst, the spectral slopes are too steep to enable
fitting the spectrum with any combination of synchrotron and SSC
emission models. However, as we showed in \S\ref{sec:fits}, Figures
\ref{fig:ee} -- \ref{fig:eB}, excellent fits are obtained using the
hybrid (thermal + non-thermal) model considered in this paper. While
we stress again that we did not make any attempt to obtain the
statistical best fits to the data, we are clearly able to obtain fits
that are within the $\pm 1 \sigma$ error bars of the data, over a
very broad band - about 6 orders of magnitude spectral range. These
fits are obtained using well understood emission mechanisms. We
can therefore conclude that, at least for this burst, the 'Band'
function is well represented by a combination of physical emission
processes. By doing so, we gain an insight into the physical
conditions in the emitting regions.

According to our interpretation, the very pronounced $\sim$~MeV peak
represent the contribution of the thermal component, and the broad
band high and low energy spectrum result from non-thermal processes.
By making this separation, we were able to constrain the hydrodynamic
properties of the outflow and the physical properties at the emission
site: the initial expansion radius was found to be $r_0 = 2.9 - 7.5
\times 10^8$~cm, the photospheric radius $r_{ph} \simeq 7.2 - 8.4
\times 10^{11}$~cm and the Lorentz factor is in the range $920 \leq
\eta \leq 1070$. The main source of uncertainty in these values is the
unknown kinetic luminosity.

Fit results from the interval during which an $11.16$~GeV photon was
observed, constrain the dissipation radius to be $r_\gamma \geq 3.5 -
4.1 \times 10^{15}$~cm. Combined measurements of the thermal and
non-thermal parts imply very high dissipation efficiency, $\epsilon_d
\simeq 85\% - 95\%$, and very high fraction of dissipated energy
carried by the energetic electrons: $\epsilon_e$ is at or above
equipartition value. Measurements of the low energy part of the
spectrum (below the thermal peak) imply high magnetic field,
$\epsilon_B$ close to equipartition. The power law index of the
electrons is more difficult to constrain, as the high energy spectral
slope results from a combination of flat (or slowly decaying)
synchrotron part, and rising Comptonized part. As such, the observed
spectral slope {\it does not directly corresponds to the electrons
  power law index}. From the numerical fits, we concluded that $2.0
\leq p \leq 2.2$ are consistent with the data, and that even $p=2.5$
is marginally consistent with the data.

The numerical results presented in Figure \ref{fig:r} show high energy
cutoff due to pair production phenomenon. While this cutoff is
consistent with the analytical prediction in equation
\ref{eq:r_gamma}, it also shows the limitation of the analytical
approximations, which are commonly in use.  Due to the inclusion of
the thermal photons, the dependence of the cutoff energy on the
emission radius is weaker than the simple approximation given in
equation \ref{eq:r_gamma}.

Additional constraints may be obtained by adding additional
information, albeit with a higher level of uncertainty. For example,
assuming the internal shocks scenario, and considering that the
$11.16$~GeV photon was seen after a delay of 11~s from the GBM
trigger, implies, for constant Lorentz factor and $\Delta \eta \simeq
\eta$, dissipation radius of $r_\gamma \sim \eta^2 c t \approx 2
\times 10^{17}$ ~cm. We point out that this assumption is consistent
with the observations, since the much more rapid variability time
($\leq 0.1$~s) can be attributed to emission from the photosphere,
rather than the high-energy non-thermal part.

While GRB090902B shows very pronounced peak and clearly separated high
energy component, which make it ideal for demonstrating our analysis
method, this is not the case in many GRBs \citep[see, e.g., recent
analysis by][]{Zhang+10}. In fact, in many GRBs, the sub-MeV peak
often seen is not as pronounced as in GRB090902B: the low and high
energy spectral slopes are not as steep, and so in many cases pure
single black body spectrum is too narrow to fit the sub MeV peak. In
some cases, e.g., GRB100724B \citep{Guiriec+10}, a weak thermal
component, that is not associated with the main peak of the spectrum,
can be identified. In many other cases, there is no clear evidence for
the existence of thermal component, as black body spectrum is not
clearly identified \citep{Nava+11}\footnote{Although the steepness of the low energy
  spectral slope can be viewed as an indirect evidence.}.  Moreover,
often the high energy spectral slope (above the thermal peak) decays
with a power law index much shallower than seen in GRB090902B. Thus,
to date, the spectrum of GRB090902B is unique by having such a
pronounced thermal peak.

Clearly, a lack of very pronounced thermal peak in most bursts is a
major drawback to the ideas raised here. There are several ways that
can explain these observations.  One possibility is suppression of the
photospheric component which is expected in Poynting flux dominated
flow \citep{ZP09, ZY11}. In this scenario, the magnetic field, rather
than the photon field, serves as an energy reservoir. As the magnetic
energy is gradually dissipated, most of the emission occurs at large
radii above the photosphere, leaving only a weak photospheric signal.
Within the fireball model itself, the pronunciation of the thermal
peak strongly depends on the value of the Lorentz factor, $\Gamma$,
which is a free parameter of the model (see equations \ref{eq:temp},
\ref{eq:luminosity}). While there is a clear indication for
$\Gamma\approx 10^3$ in GRB090902B and several other bursts, in many
bursts the value of $\Gamma$ is lower, $\sim 10^{2.5}$ \citep[see,
e.g.,][]{Racusin+11}. Thus, for these bursts, the thermal emission,
while expected to exist, is not as pronounced as in GRB090902B.

Alternatively, these observations can be explained in a framework
similar to the one used here, namely that the spectrum may be
dominated by Comptonization of the thermal photons, following energy
dissipation that occurs close to the photosphere. As shown by
\citet{PMR05, PMR06}, in this scenario, multiple Compton scattering by
electrons at a quasi steady state produces a flat energy spectra for a
large parameter space region. Since the dissipation radius
($r_\gamma$) can in principle take any value above the saturation
radius, different dissipation radii can lead to very different
observed spectra \citep[see][]{PMR06}; in particular, high energy
power law tail is obtained \citep{LB10}. Thus, sub-photospheric
dissipation can reproduce the Band function \citep{Ryde+11}. 
In fact, it could very well be that the uniqueness of
GRB090902B originate from a very large dissipation radius, $r_\gamma \gg
r_{ph}$. Only under this condition one is able to make such a clear separation
between the thermal and non-thermal parts of the spectrum, which are
otherwise coupled.

This possibility can also explain the lack of GeV emission in many
bursts. If $r_\gamma$ is not much larger than $r_{ph}$, then, by
definition, the optical depth to scattering is high. Since the cross
section to pair production is similar to $\sigma_T$, the optical depth
to pair production is high too (see equation \ref{eq:r_gamma}). As a
result, GeV emission is attenuated.  Alternatively, attenuation of GeV
emission is expected if the outflow is highly magnetized: in this
scenario, synchrotron emission dominates over IC scattering, leading
to attenuation of the high energy emission.

In addition to the complex relations between the thermal and
non-thermal parts of the spectra, there are two effects which are
often being neglected. First, the fits are often made to {\it time
  integrated} spectra. As the properties of the outflow, in particular
the Lorentz factor, vary on a very short duration (variation can be
expected on time scale of the order of $r_0/c$, i.e., $\mathcal{O}(10
{\rm ms})$), the black body spectrum is often smeared. Even more than
that, as was shown in \citet{Peer08} and \citet{PR10}, at any given
instance, an observer sees simultaneously thermal photons emitted from
different radii and different angles to the line of sight, hence
having different Doppler shifts. As a result, the expected
photospheric emission is {\it not} a pure black body, but a
combination of black body spectra with different amplitudes. Thus, in
fact, one expects to see a multi-color black body, as is indeed
seen. The full theory of multi-color black body emission from
relativistically expanding plasmas recently appeared in \citet{PR10}.

The results of the fits to GRB090902B imply a very high value of
$\epsilon_e$, close to or even above equipartition, and an even higher
value of $\epsilon_d$, 85\%- 95\%. These values are much higher than
the values predicted by the internal shocks model: the typical
efficiency in energy dissipation by internal shocks is no more than a
few percent. The results obtain here (note that the exact nature of
the dissipation process is not specified) thus raise another issue as
to the validity of the internal shock model scenario. This is added to
GRB080916C, in which detailed analysis by \citet{ZP09} concluded that
an additional source of energy must exist between the photospheric
radius and the dissipation radius. The most plausible source of energy
considered is magnetic, i.e., a Poynting dominated outflow
\citep[see][]{ZY11}. We can conclude that the high efficiency
required, may hint toward Poynting dominated outflow in the case of
GRB090902B as well, and may even be a general requirement for all the
bursts with pronounced high energy emission observed by {LAT}. On the
other hand, even if the outflow in GRB090902B is Poynting flux
dominated, we do not expect too high ratio of the Poynting to kinetic
luminosity, $\sigma$, since $\sigma \gg 1$ results in suppression of
the photospheric emission, which is not observed in this burst.  A
more generalized treatment of photospheric models with arbitrary
magnetization is outside the scope of this manuscript, and is left for
future work.

The separation made in this work into two emission zones, namely
thermal emission originating from the photosphere, and non-thermal
emission originating from energy dissipation at larger radii, provides
a natural explanation to the delay of the high energy photons, often
seen in {Fermi-LAT} bursts \citep{Abdo09, Ryde+10, TWM11, GGNC10}. In
our model, the non-thermal photons originate from dissipation above
the photosphere, hence they are naturally seen at a delay with respect
to the photospheric (thermal) photons, which are always the first to
be observed.  A pronounced thermal component at early times also
provides a natural explanation to the harder slope at low energy
observed during the first 1-2 s in many long bursts
\citep{Ghirlanda+09}.  In this general framework, the origin of the
high energy, non-thermal photons can also be hadronic, as recently
suggested \citep{RDF10, Asano+10}. Nonetheless, we showed here that a
leptonic origin is consistent with the data. Our model has the
advantage that it does not require a large amount of energy in the
hadronic component. Moreover, we did not specify the origin of the
dissipation that lead to the emission of the high energy photons.
Recently, there were several suggestions of external shock origin of
these photons \citep{KD09,KD10,GGNC10}. However, by fitting the broad
band data we showed that excellent fits can be obtained if both the
characteristic synchrotron breaks are below the {\it Fermi} energy
band, i.e., less than $\sim 10 \keV$. Thus, our model is consistent
with internal dissipation origin of the non-thermal spectrum
\citep[see further discussion in][]{PN10,Zhang+10}.

\vskip 4mm
{\bf ACKNOWLEDGEMENTS}
\vskip 3mm
This research was supported by the Riccardo Giacconi Fellowship award
of the Space Telescope Science Institute. FR acknowledges financial
support by the Swedish National Space Board. AP wishes to thank Dale
Frail, Jeremy Schnittman, Andy Fruchter, Kuntal Misra, Zeljka Bosnjak,
Mario Livio and Kailash Sahu for useful discussions.

\appendix
\section{Decomposition of the spectrum into its basic physical
  ingredients} 
\label{sec:decomposition}

The observed spectrum results from synchrotron emission, SSC and
thermalization of the thermal photons, with a high energy cutoff
resulting from pair production. As discussed in \S\ref{sec:results},
the three emission processes are expected to have similar
contributions to the observed spectrum of GRB090902B. Therefore, a
``clean'' decomposition into the spectral ingredients, as is presented
in, e.g., \citet{SE01}, does not exist in practice. 

The high energy spectral slope (above the thermal peak) does not have
a direct correspondence to the power law index of the accelerated
electrons. Therefore, a decomposition of the spectrum into the basic
radiative processes can be useful in understanding the physical
processes that shape the spectrum.  Fortunately, such a decomposition
can (up to some level) be done numerically.

\begin{figure}
\begin{center}
\rotatebox{0}{\resizebox{!}{60mm}{\includegraphics{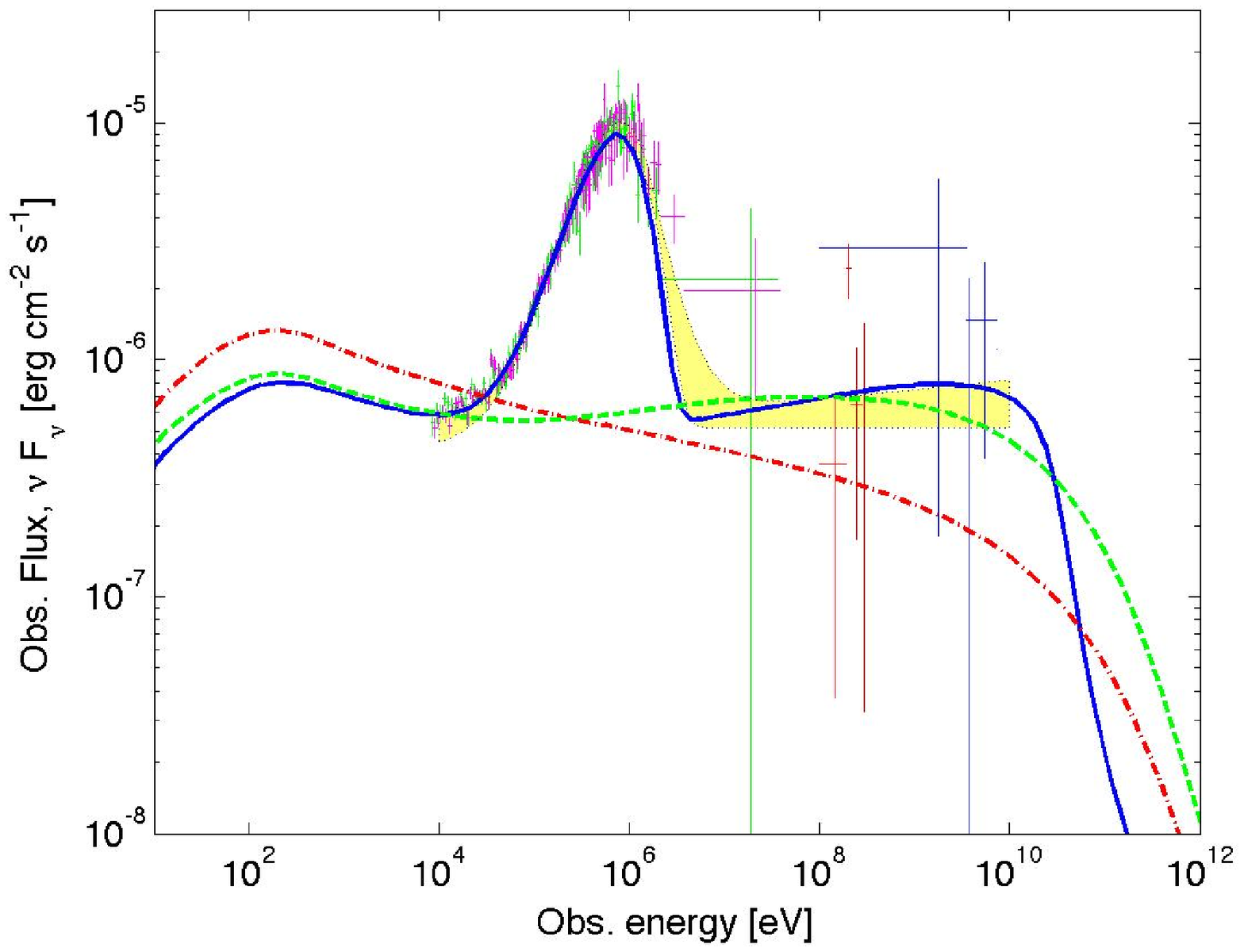}}}
\caption{Demonstration of spectral decomposition into basic physical
  ingredients. The dash-dotted (red) curve show the spectrum that
  would have obtained if synchrotron radiation was the only source of
  emission. The dashed (green) curve show the resulting spectrum from
  synchrotron and SSC, and the solid (blue) curve show the spectrum
  with the full radiative ingredients (synchrotron, SSC, the thermal
  peak at $\sim \MeV$, and
  Comptonization of the thermal photons). Dissipation radius
  $r_{\gamma} = 10^{16} \cm$, , $\epsilon_e = 0.5$, $\epsilon_B =
  0.33$, $p=2.2$ and all other parameter values same as in Figure 1 are
  chosen. The low energy spectral slope (below the thermal peak) is
  mainly due to synchrotron emission, and is thus sensitive to the
  power law index of the accelerated electrons. However, the high
  energy part (above the thermal peak) results from all of the
  radiative processes, and therefore cannot be used directly to
  constrain the values of the free model parameters.}
\label{fig:decomposition}
\end{center}
\end{figure}

The numerical results are presented in Figure \ref{fig:decomposition},
for dissipation radius $r_{\gamma} = 10^{16} \cm$, and power law index
$p = 2.2$. The dash-dotted (red) curve represents the spectrum that
would have been obtained if only synchrotron emission was
considered. As the cooling frequency is low, the spectral index in the
entire {\it Fermi} range is $\nu F_{\nu} \propto \nu^{1-p/2} \propto
\nu^{-0.1}$. When SSC is added (dashed, green curve) the spectral
slope at low energies (below $\approx 10^4 $~eV) is not affected,
while the spectrum at high energies becomes nearly flat. This results
from a combination of a decreasing synchrotron flux and an
increasing SSC flux.  Note that in this scenario, the flux at low
energies is weaker than the flux obtained for the pure synchrotron
scenario, due to the fact that part of the electron energy is
converted to SSC. Finally, the solid (blue) curve shows the combined
effects when all the physical ingredients are added. Inclusion of the
thermal photons does not affect the low energy spectral slope (below
the thermal peak). However, Comptonization of the thermal photons
contribute to the more pronounced spectrum at high energies. In
addition, the inclusion of thermal photons leads to a sharper high
energy cutoff, resulting from pair production.

\end{document}